\documentclass[12pt]{article}
\usepackage{graphicx}
\usepackage{epsfig}

\usepackage{amssymb}

\begin{document}

\title{ On the Zipf strategy for short-term investments in WIG20 futures}
\author{Bartosz Bieda$^{(1)}$, Pawe{\l} Chodorowski$^{(2,3)}$ and Dariusz Grech$^{(3)}$\footnote{dgrech@ift.uni.wroc.pl}}
\date{}

\maketitle
\begin{center}
(1)Institute of Telecommunication, Teleinformatics and Acoustics, Wroc{\l}aw University of Technology, Wybrze{\.z}e St. Wyspia{\`n}skiego 27, PL-50-370 Wroc{\l}aw, Poland\\
(2) Bank Zachodni WBK SA (BZWBK), Finance Division, Rynek 9/11,\\ PL-50-950 Wroc{\l}aw, Poland\\
(3) Institute of Theoretical Physics, University of Wroc{\l}aw,\\ Pl. M.Borna 9, PL-50-204 Wroc{\l}aw, Poland
\end{center}

\hspace{1.5 cm}

\begin{abstract}
We apply the Zipf power law to financial time series of WIG$20$ index daily changes (open-close). Thanks to the mapping of time series signal into the sequence of $2k+1$ 'spin-like' states, where $k=0, 1/2, 1, 3/2, ...$, we are able to describe any time series increments, with almost arbitrary accuracy, as the one of such 'spin-like' states. This procedure leads in the simplest non-trivial case $(k = 1/2)$ to the binary data projection. More sophisticated projections are also possible and mentioned in the article. The introduced formalism allows then to use Zipf power law to describe the intrinsic structure of time series. The fast algorithm for this implementation was constructed by us within $Matlab^{TM}$ software. The method, called Zipf strategy, is then applied in the simplest case $k = 1/2$ to WIG 20 open and close daily data to make short-term predictions for forthcoming index changes. The results of forecast effectiveness are presented with respect to different time window sizes and partition divisions (word lengths in Zipf language). Finally, the various investment strategies improving ROI (return of investment) for WIG20 futures are proposed. We show that the Zipf strategy is the appropriate and very effective tool to make short-term predictions and therefore, to evaluate short-term investments on the basis of historical stock index data. Our findings support also the existence of long memory in financial data, exceeding the known in literature  $3$ days span limit.
\end{abstract}
$$
$$
\textbf{Keywords}: Zipf law, econophysics, investment strategy, long memory, time series, complex systems, futures contracts\\

\textbf{PACS:} 05.45.Tp, 89.75.Da, 05.40.-a, 89.75.Da, 89.65.Gh

\section{Introduction}

The Zipf law has originally been introduced in linguistic [1] to describe the frequency occurrence of different words in written text. Since then, the similar law has been observed in systems of various origin and in many disciplines of science, economy, finances, biology, sociology, medicine, physiology and many others [2]. In the general formulation one may describe the Zipf law as follows. Let $\{e_1, e_2, e_3, ..., e_n\}$ be an arbitrary system of the countable number of events ordered in such a way that the frequency $f_k$ of the event $e_k$ is bigger than the corresponding frequency $f_{k+1}$ of $e_{k+1}$ $(k=1,2,...,n-1)$. We say that the Zipf power law is satisfied in this system of events if for normalized frequencies $f_k$ there exists a real number $\zeta>0$ (called the Zipf exponent) such that:

\begin{equation}
f_k \sim k^{-\zeta}
\end{equation}
The index $k$ is then called the rank of event $e_k$.\\
 The origin of Zipf law is not well understood. Nevertheless, we know since the paper by Czirok {\em et.al} [3] that if correlations exist in the complex system then the frequencies of various events in this system obey the Zipf law. The conjecture between the Hurst exponent $H$ [4], responsible for the level of correlations (autocorrelations) in the system, and the Zipf exponent $\zeta$ is given by [3,5]:

\begin{equation}
\zeta = |2H-1|
\end{equation}

The inverse statement is not true [6] -- the existence of Zipf law in the system does not imply automatically correlations nor long memory between events in the system. However, if for shuffled data (the shuffling procedure may look differently for various considered systems of data) the Zipf law appears with Zipf exponent $0<\zeta_{shuff}<\zeta_{r}$ where $\zeta_{r}$ was calculated for the original data, then we may conclude that
the power law acting in given system comes as a result of
correlations between different events and $\zeta_{shuff}$ describes the bias level of the system [7].
Thanks to the mapping between time series signal and letters [7,8], one may investigate the interior structure of time series with the use of Zipf power law. The generalization of such mapping can be made as follows.\\
Let $x_1, x_2, ..., x_n, x_{n+1}$ is the given discrete time series with increments $\Delta x_i = x_{i+1}-x_i$ $(i=1, 2, ..., n)$. The sequence $\{\Delta x_i\}$ may be mapped into the n-string of $2k+1$ states with integer and positive $2k$ ('spin-like' chain). In the simplest case ($k=1/2$), we have only two possible states: $u$ (up) and $d$ (down), corresponding to $\Delta x_i>0$ or $\Delta x_i<0$ respectively (the case $k=0$ is trivial).\\
 Introducing the threshold $l>0$, we may consider three admissible states: $\Delta x_i>l$, $\Delta x_i<-l$ or $-l\leq \Delta x_i\leq l$, denoted as $u$, $d$, $s$ ($s$ for 'stabile') respectively. The whole range of thresholds may be introduced this way, leading to more sophisticated discrete approximation of time series increments. This mapping may be done with arbitrary accuracy if the number of thresholds (states) is sufficiently large. However, one has to remember that if the number of admissible states increases, we need also larger amount of data (the time series length) to be able to apply the Zipf law for sufficiently large statistics and to overcome the bias resulting from variety of words put into the relatively short text. This might be a problem in practical applications due to the finite and usually short real time series data, e.g. in finance. Therefore we consider two states further on in this paper.

\section{Zipf strategy description}

The prediction for future behavior of time series based on the Zipf law can be summarized in few steps. First, we translate the given time series into the sequence of $u$, $d$ letters (text). Unfortunately, no words are distinguished so far in this text. Therefore one should divide it  into non-overlapping subsets containing $m$ letters each. We call them 'words' of length $m$ or shortly '$m$-words'. Next, the window length $w$ (amount of data) over which the Zipf analysis will be performed, must be chosen.  We are going to make the local Zipf analysis what means that this window is shifted one session forward each trading day. Thus, we always deal with  $w$ available data back in time. We have chosen  $w=400, 500, 600, 700, 800$ in our analysis.\\
 The crucial role is played by $m$-words. They reveal the structure and eventual possible memory in financial data. Once we want to make prediction for the sign of time series change in one trading day ahead, we construct the $m$-word as $(l_1,l_2,...,l_{m-1},x_m)$, where $l_1,...,l_{m-1}$ are known and describe the known evolution of time series in last $m-1$ days, while $x_m$ is the unknown behavior, still to be predicted. There are two possibilities for $x_m =u,d$ in two states scenario. Let $p_u=p(l_1,l_2,...,l_{m-1},u)$ is the normalized frequency of $(l_1,l_2,...,l_{m-1},u)$ appearance in the $w$-length text. The corresponding notation for $p_d$ follows. One obtains from the Zipf law

\begin{equation}
\frac{p_u}{p_d} \sim {(\frac{R_u}{R_d}})^{-\zeta}
\end{equation}
where $R_{u(d)}$ are respective ranks for $m$-words in $w$-length text. The additional constrain $p_u+p_d=1$ leads from Eq.(3) to the solution:

\begin{equation}
p_u={R_u}^{-\zeta}/({R_u}^{-\zeta}+{R_d}^{-\zeta}), \qquad p_d={R_d}^{-\zeta}/({R_u}^{-\zeta}+{R_d}^{-\zeta})
\end{equation}
If $p_u>p_d$, one gets a signal that the index is very likely to increase in the $m$-th session, otherwise ($p_u<p_d$), the index is indicated to fall down.
In the proposed strategy we may open the so called short or long position. It means that we can  sell or buy futures contracts at the opening price each trading day. Then, at the end of a trading day, we close this position (we buy or sell respectively) at the closing price. In order to reduce transaction costs we decided to take the day
trading, i.e. open and close positions on the same trading day. It saves about $1/3$ cost of a commission charge depending on the offered tariff charge. This is why instead of closing prices each day, we are more interested in open $-$ close WIG20 values each trading day. Fig.1a represents the history of WIG20 closing prices in the period: December 20 '99 - May 25 '10  and the corresponding history of cumulative daily changes in Fig.1b (i.e. the difference between closing and opening price of the index each trading day, summed over the running period). The latter index reflects changes on the market that take place only during the transaction day. It neglects changes made during the night, mostly affected by overseas (US) trading. We will call such index 'the day-light' WIG20 (WIG$^{dl}20$).\\
The proposed strategy can obviously be expanded to more than one day prediction. In such a case, the considered $m$-word would be $(l_1,l_2,...,l_{m-2},x_{m-1},x_m)$ for two days prediction with unknown string of letters $x_{m-1}$, $x_m$. Similarly, three days prognosis may be done. These strategies would correspond to futures contracts started particular day in the morning and effective one or two days ahead after closing the session.\\
Let us recall also few basic information on futures contracts we will use.\\
Futures contracts are example of derivatives.
In case of derivatives, the words "buy" and "sales" are not used, because in some cases it may be misleading. Instead, the terms "long position", i.e. an obligation to buy the underlying asset at a fixed price and "short position", i.e. an obligation to sell the underlying asset at a fixed price, are used. Thus, the meaning of "short" and "long" has no relation to the time a position is being hold. Unlike options, futures contracts are symmetrical. It means that both parties are obligated to provide or buy the underlying asset. There is no cash flow between the parties at the opening of the contract. The crucial role is played in futures contracts by the so called margin. It is an initial deposit we have to make while opening the position. It is a hedge against the risk of default of the contract. Its minimum amount is determined by the clearing house. The margin is only a part of the contract value and it creates an important financial leverage, which amplifies both gains and losses.
The margin is always required. If the WIG20  changes by one point, it results in a gain or loss of PLN $10$ for one contract. The margin is about $10\%$ and it depends mainly on the volatility of the index.
 While opening the position, the corresponding margin is blocked in the account. At the end of the day return is calculated (this process is called marking-to-market). The profits are added to the deposit, and the loss is subtracted. If the amount of deposit falls below a certain minimum value, known as maintainance deposit, trader will get a margin call to its initial value. If he does not do this, his position will be automatically closed.\\
Let us look at the following example explaining the leverage role of futures contracts.
Let us assume that the WIG20 index amounts to $2 500$ points and the required margin is $10\%$. The value of one contract is then PLN $25,000$ ($2500$ points $\times$ PLN $10$). Assume we open long position in two contracts. Thus the initial margin amounts to PLN $5,000$. If WIG20 increases up to $2 550$ points at the end of a trading day and we close the position then the value of one contract will be PLN $25,500$. So we would receive the profit of PLN $1,000$  what makes $20\%$ of our initial investment. At the same time  WIG20 has increased just $2\%$.

\section{Results and Conclusions}

First we have checked the performance of the Zipf law locally, for various window lengths $w$ and for different word lengths $m$. The used $w$ values cover the whole period of available data for WIG20 and change between $\sim 1.5$ up to $\sim 3$ years of data ($w=400\div 800$). The examples of fit for the Zipf power law in logarithmic scale for these parameters are shown in Figs.2,3,4. All plots represent two fits, made correspondingly for the original and shuffled data (WIG$^{dl}20$ changes), translated into binary sequence and then divided into $m$-word pieces. The scaling regime, although short due to respectively small number of data $w$, enabled to find the local Zipf exponent $\zeta_r$ for the original, and $\zeta_{shuff}$ for shuffled, uncorrelated data. We have observed in majority of cases that $\zeta_{r}>\zeta_{shuff}$\footnote{this relation was sometimes violated for small $w$ where the bias of artificial autocorrelations emerging from small statistics is present even for uncorrelated data}. Due to lack of space, only nine exemplary plots are shown in Figs.2,3,4 as the illustration of this phenomenon.

\begin{table}
\begin{center}

\begin{tabular}{| c | c | c || c | c || c | c |}
\hline
   & \multicolumn{6}{|c|}{Word length $m$} \\
 $w$& \multicolumn{2}{|c}{4} & \multicolumn{2}{c}{5} & \multicolumn{2}{c|}{6} \\ \hline

    & Accuracy & Profit (PLN) & Accuracy & Profit (PLN) & Accuracy & Profit (PLN) \\ \hline
400 & 53.0\% & 4 570 & 53.0\% & 7 630 & 55.4\% & 13 310 \\ \hline
\textbf{500} & \textbf{53.6\%} & \textbf{6 230} & \textbf{54.6\%} & \textbf{9 030} & \textbf{57.0\%} & \textbf{19 230} \\ \hline
600 & 53.0\% & 6 070 & 54.4\% & -190 & 55.2\% & 16 010 \\ \hline
700 & 53.0\% & 3 570 & 54.2\% & 5 370 & 55.2\% & 20 130 \\ \hline
800 & 53.6\% & 8 390 & 53.8\% & 1 690 & 54.8\% & 15 490 \\ \hline

\end{tabular}
\caption{Results for the local Zipf strategy applied to WIG20 in the period May 8'08-May 25'10}
\end{center}
\end{table}

The detailed dependence of $\zeta_r$ on the number of data taken to the analysis and on the length $m$ of the probe word can also be found (see Fig.5 for details).
It is evident from this figures that for shorter window lengths $w<300$, corresponding to no more than one year of trading, the bias of artificial autocorrelations is more visible due to insufficient data statistics. On the other hand, for longer $w$ exceeding 3 years (more than $800$ sessions), the local Zipf exponent  approaches values $\zeta_{shuff}\lesssim 0.15$ shown for shuffled data in Figs.2,3,4. Such behavior may be explained as the effect of autocorrelation decay in global long-term data. This situation is somehow similar to the case with local Hurst exponent being estimated for too short or too long time windows [9]. The proper choice of moving window length $w$ was crucial for the local Hurst exponent estimation to eliminate the bias and to extract the possible autocorrelation signal [10,11]. We decided to use $400<w<800$, i.e. the middle part of $w$ spread from the one shown in Fig.5.\\
The results for $\zeta_r, R_u, R_d$ calculated in subsequent moving windows like in Figs.2,3,4, allow to predict $p_u$ vs $p_d$ frequencies in the local Zipf strategy based on Eq.(4). The final outcomes of this strategy is collected in Table 1 for $m=4,5,6$ and $w=400, 500, 600, 700, 800$ respectively.

 More than $50\%$ of WIG$^{dl}20$ changes is well predicted in all considered cases. The best performance is found for $w=500$ (marked in bold font in Table 1), in particular for $m=6$. It figures the return profit around $870\%$ in two years (May'08-May'10) since the average initial deposit for WIG20 futures in that period was calculated as  PLN $2,217$ assuming the $10\%$ margin . The performance for other $m$-words in this period is also plotted in Fig.6.
 Let us note that the longer investment horizon the better results of the strategy are achieved thanks to larger statistics.\\

Our findings support the statement that Zipf law can be used as the marker of long memory effects in financial data [7, 12] and the useful basis for investment strategy in futures contracts. Moreover, the comparison of Zipf exponents for ordinary and shuffled financial data indicates that this analysis is more sensitive than the observation of heavy tails in return probability density function. It is well known that heavy tails in such distributions decay for time-lags yet above $3$ trading days [13]. We were able to see the memory effect for time-lags exceeding this level (see Fig.4.) and we confirm the $6$ day memory effect in financial data suggested for particular two stocks: SGP from NASDAQ and OXHP from NYSE already in [12]. Our results indicate this might be a more general property for financial time series.\\
 The next important remark is that the local Zipf strategy gives excellent results when applied to futures contracts independently on the current trend on the market. This strategy, although applied in this paper to WIG20 data only, can be extended to other financial, FOREX or commodity markets data where futures contracts are available. The strategy offers also possibility to make an automatic numerical application, what might be important for practitioners and financial analysts as the new indicator of technical analysis.
\\
\\
\\

%%%%%%%%%%%%%%%%%%%%%%%%%%%%%%%%%%%%%%%%%%%%%%%%%%%%%%%%%%%%%%%%%%%%%%%%%%%%%%%%%%%%%%%%%%%%%%%%%%%

\newpage

\begin{figure}
\begin{center}
{\psfig{file=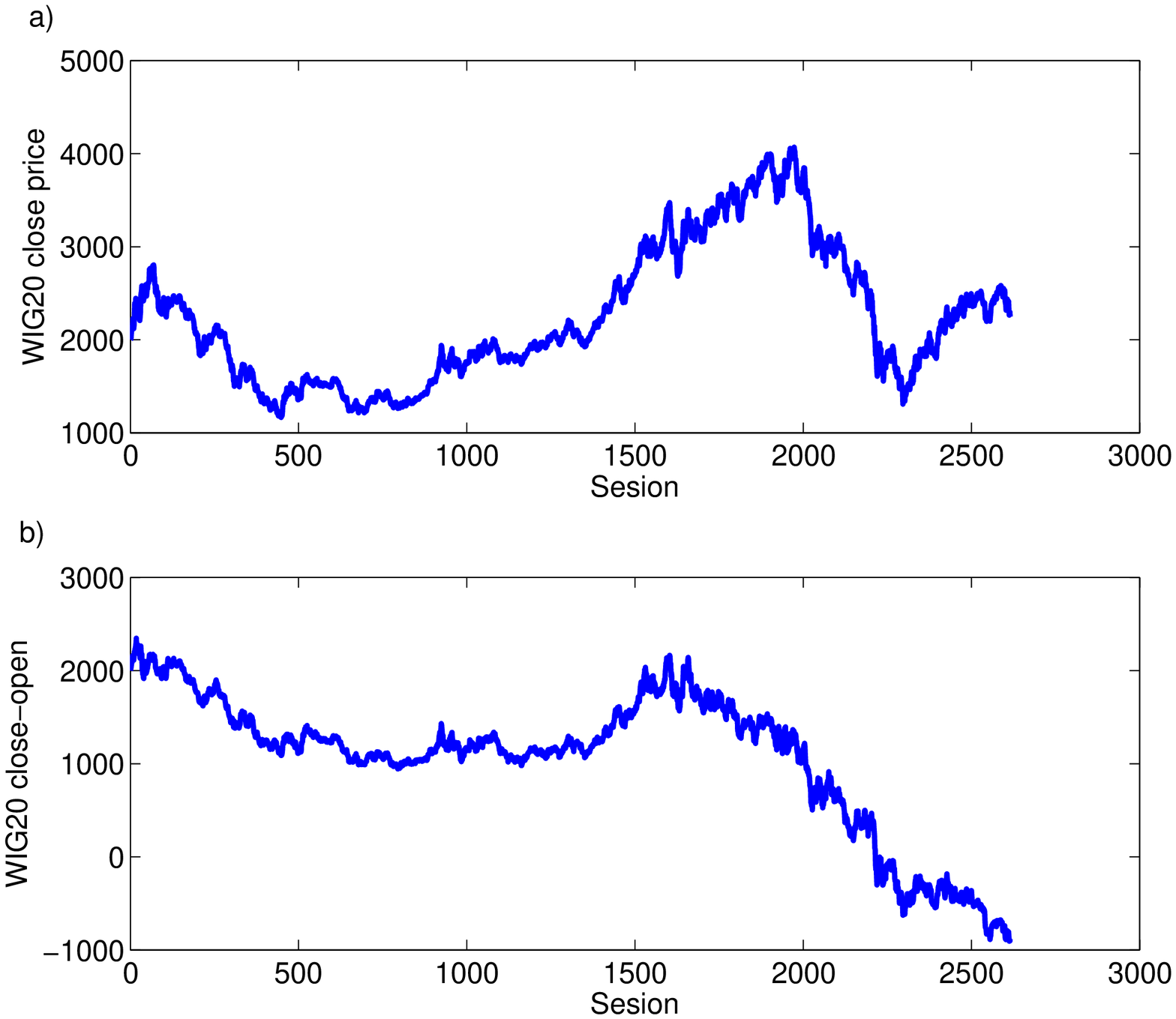,width=17cm,angle=0}}
\end{center}
\caption{The closure day WIG20 index time history Dec.20'99--May 25'10 (a) and the corresponding  artificial "day-light" WIG$^{dl}20$ index (b) in the same period.
}

\end{figure}

\begin{figure}
\begin{center}
{\psfig{file=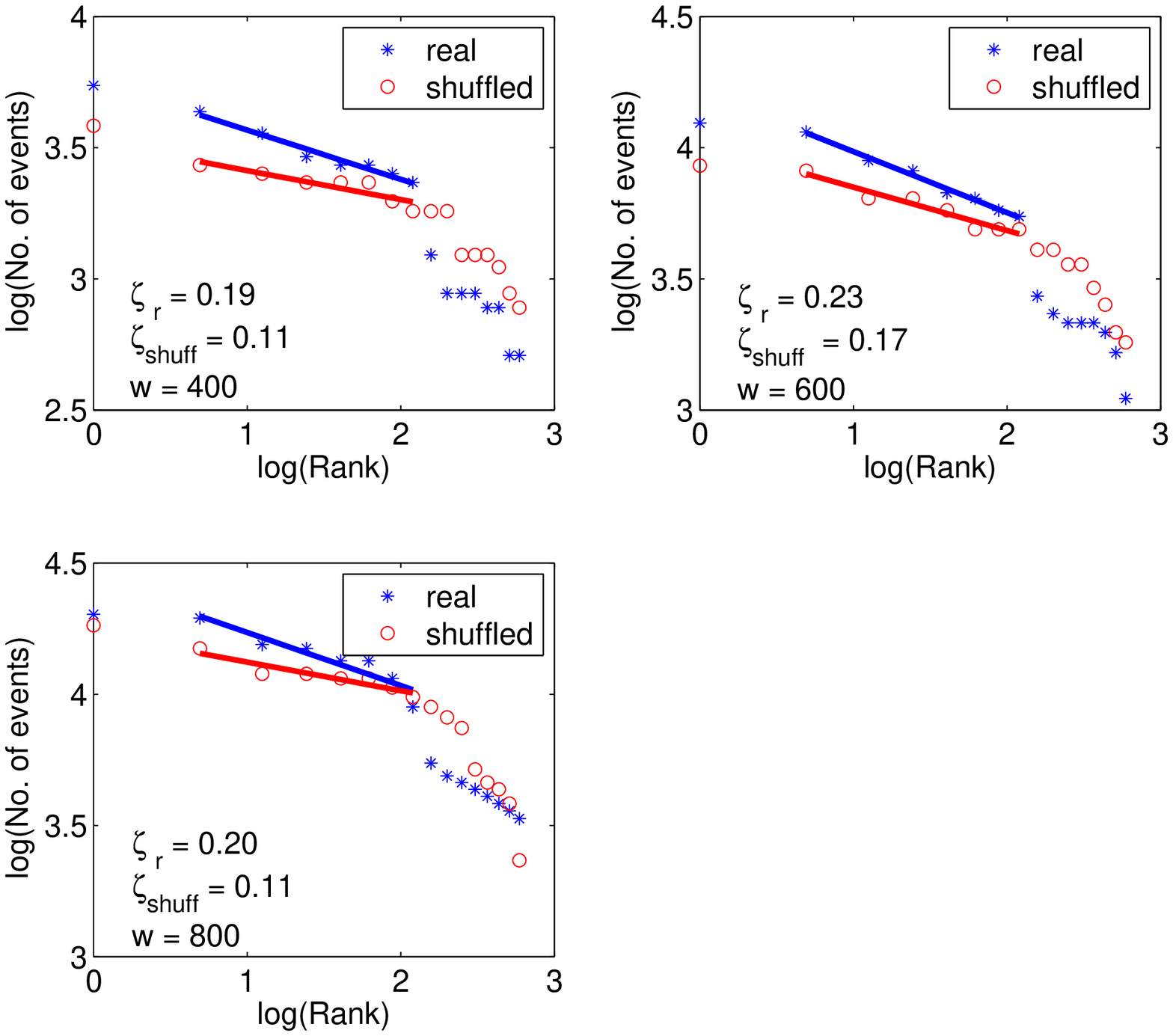,width=17cm,angle=0}}
\end{center}
\caption{ Examples of Zipf analysis for real (stars) and shuffled (circles) data from WIG$^{dl} 20$ index of length $w=400, 600, 800$ respectively, divided into words of length $m=4$ trading days.}
\end{figure}

\begin{figure}
\begin{center}
{\psfig{file=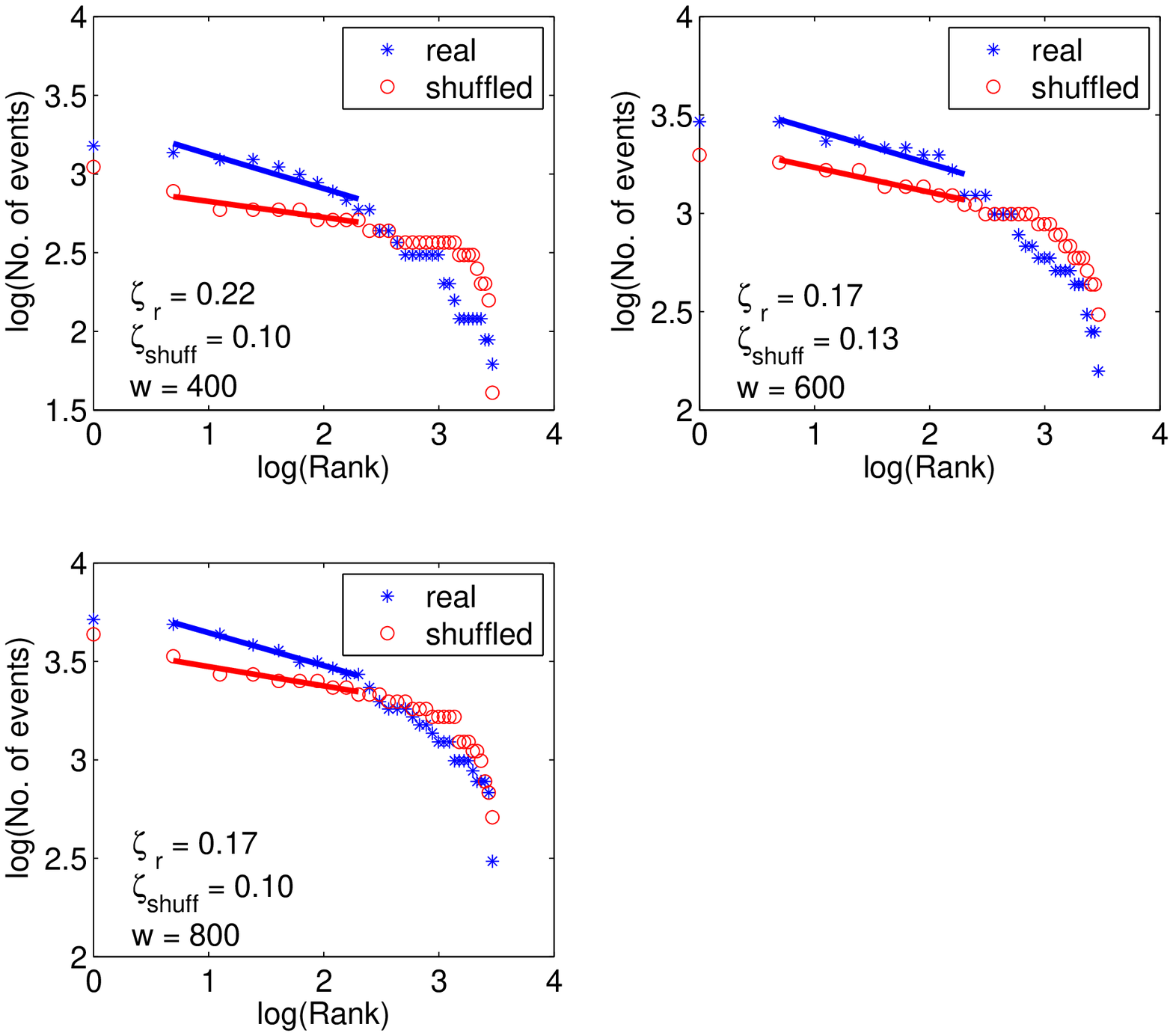,width=17cm,angle=0}}
\end{center}
\caption{ Examples of Zipf analysis for real (stars) and shuffled (circles) data from WIG$^{dl} 20$ index of length $w=400, 600, 800$ respectively divided into words of length $m=5$ trading days.}
\end{figure}

\begin{figure}
\begin{center}
{\psfig{file=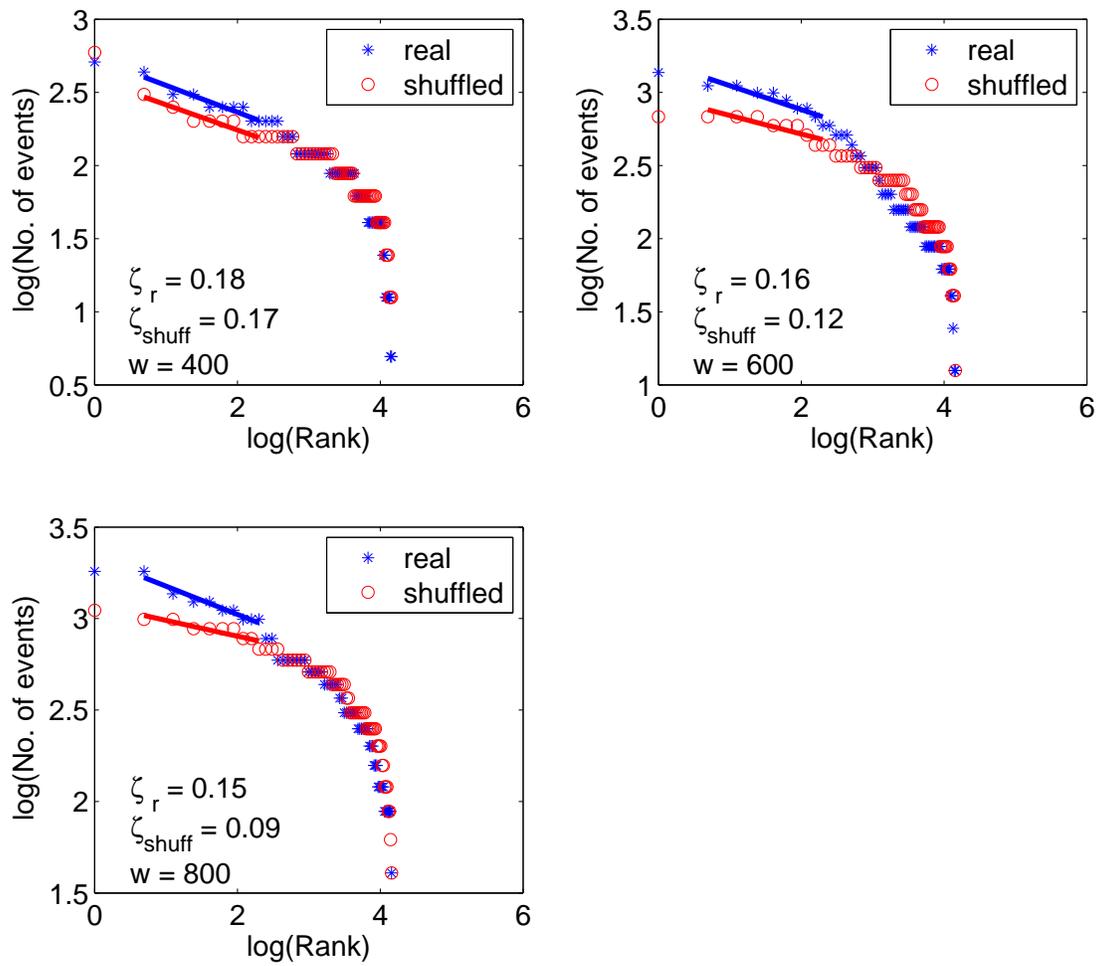,width=17cm,angle=0}}
\end{center}
\caption{Same as in Figs.2,3 but for $m=6$ $m$-words. }
\end{figure}

\begin{figure}
\begin{center}
{\psfig{file=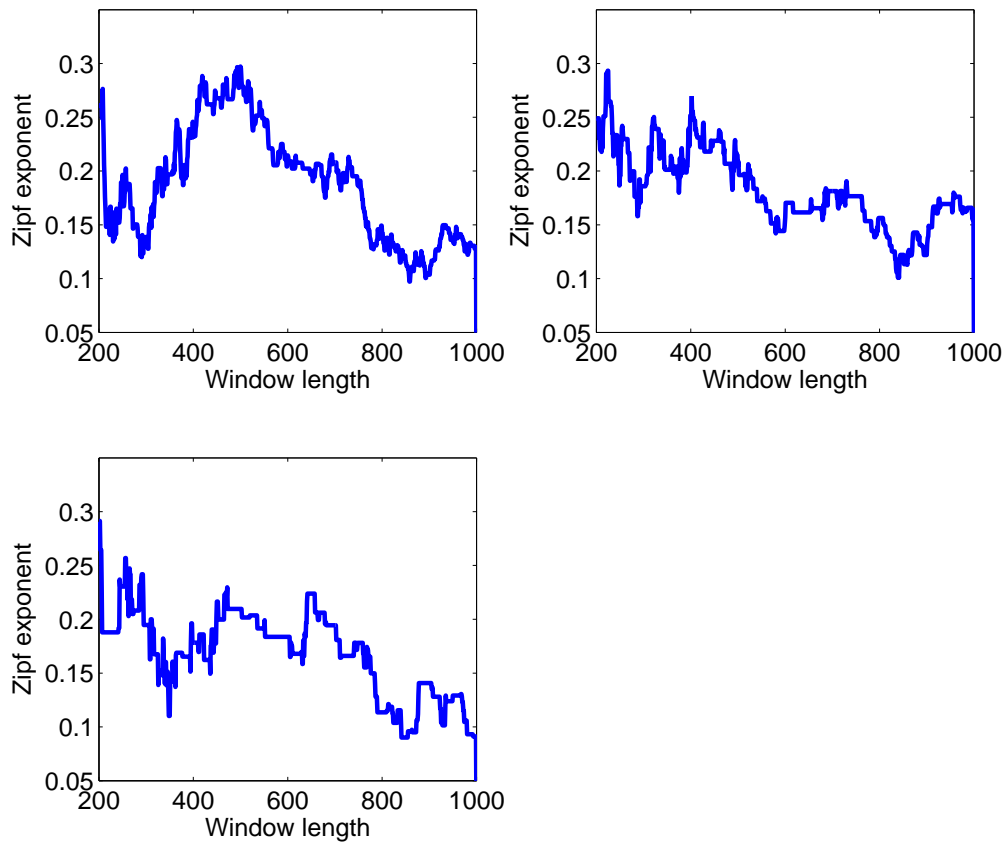,width=15cm,angle=0}}
\end{center}
\caption{Evolution of local Zipf exponent for particular choice of word length $m=4$ (top left), $m=5$ (top right) and $m=6$ (bottom) as the function of time window length $w$.}
\end{figure}

\begin{figure}
\begin{center}
{\psfig{file=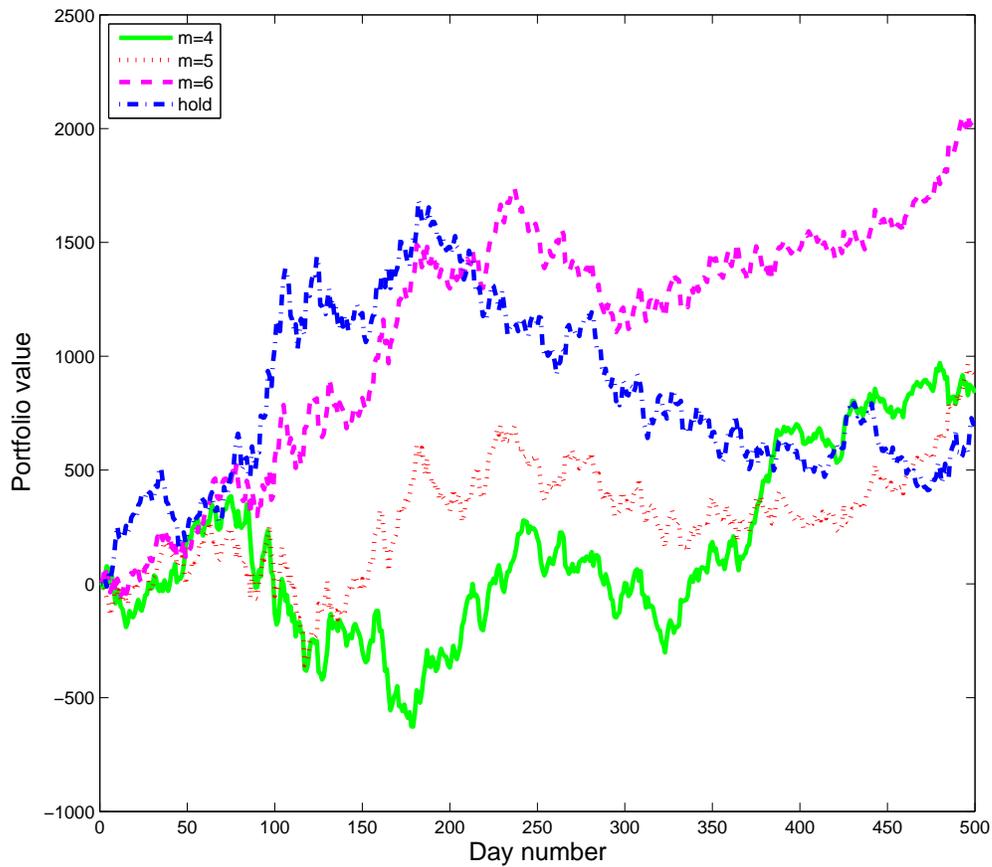,width=15cm,angle=0}}
\end{center}
\caption{Results of Zipf strategy with word length $m=4$, $m=5$ and $m=6$ applied for WIG$^{dl}20$ data for $500$ consecutive trading days starting from May 8'08 till May 25'10 }
\end{figure}
\end{document}